%% file: main.tex
\documentclass[a4paper]{article}

\usepackage[utf8]{inputenc}

\usepackage[T1]{fontenc}
\usepackage{subfiles}
\usepackage{url}
\usepackage{physics}
\usepackage{braket}
\usepackage[official]{eurosym}
\usepackage{geometry}
\usepackage{authblk}
\usepackage{caption}
\usepackage{subcaption}
\usepackage{verbatim}
\usepackage[table,xcdraw]{xcolor}
\usepackage{enumitem, kantlipsum}
\usepackage{fancyvrb}
\usepackage{longtable}
\usepackage{lscape}
\usepackage{amssymb,amsmath}
\usepackage{graphicx}
\usepackage{float}
\usepackage{multirow}
\usepackage{tabularx}
\usepackage{marginnote}
\usepackage{amsmath}
\usepackage{flexisym}
\usepackage{amsmath}
\usepackage{xparse}
\usepackage{listings}
\usepackage{tikz}
\usepackage{upquote}
\usepackage{gensymb}
\title{
Quantum Computing - A new scientific revolution in the making
}

\author[1,5]{K.~Bertels}
\author[4,5]{E.~Turki}
\author[3,5]{A.~Sarkar}
\author[2,5]{I.~Ashraf}
\author[1,5]{T.~Sarac}

\affil[1]{University of Ghent, Belgium}
\affil[2]{Delft University of Technology, Netherlands}
\affil[3]{HITEC University, Pakistan}
\affil[4]{University of Normandy-Rouen, France}
\affil[5]{All authors collaborated for many years resulting on the PISQ idea as explained in this paper}
\begin{document}

\maketitle

\begin{abstract}
After spending 10 years in Quantum Computing (QC) as computer engineers, space and chemical engineers, and given the impeding timeline of developing good quality quantum processing units, it is the moment to rethink the approach to advance quantum computing research.  
Rather than waiting for quantum hardware technologies to mature, we need to start assessing in tandem the impact of the occurrence of quantum computing, or rather Quantum Computing Logic (QC-Logic) on various scientific fields.  This is where the subtitle comes from. A new scientific revolution is unfolding.
Within the context of making real scientific progress, we need to use an additional and  complementary  approach than proposed by the NISQ-program or any follow-up approach.   We have to be aware of the fact that defining, implementing and testing the quantum concepts for any field is a tremendous work. 
The main reason is that QC initiates an overall revolution in all scientific fields, and the way those machines will be used in daily life is a very big challenge.  That is why we propose a complete update of the first PISQ-paper. Evidently, we still advocate the additional \textbf{ PISQ-approach}: Perfect Intermediate-Scale Quantum computing based on a well established concept of perfect qubits. We  expand the quantum road map with \textbf{(N)FTQC}, which stands for (Non) Fault Tolerant Quantum Computing.
This will allow researchers to focus  exclusively on the development of new applications by defining the algorithms in terms of perfect qubits and evaluate them in two ways. Either executed on quantum computing simulators that are executed on supercomputers, or apply hardware-based qubit chips. This approach will be explained in this paper.
Our planet needs a long-term vision and solution.  It will enable universities and companies, alike, to accelerate the development of new quantum algorithms, build the necessary know-how and thus addressing one of the key bottlenecks, within the quantum industry, which is the lack of talents to develop well-tested quantum applications.

\end{abstract}


\input{01-Introduction}

\input{02-FullStack}
\input{02-Problems}
\input{03-00-QRoadMap}

\input{04-00-QExamples.tex}
\input{05-PathForward}

\input{06-Conclusion}

\section*{Authors}

\begin{description}
    \item[Koen Bertels] is  part-time full Professor at the University of  Ghent where he teaches  Quantum Computer Architecture.  While at Delft University of Technology, he worked on the definition of the full stack and was in close collaboration with INTEL.  He now focuses on the use of perfect qubits to realise a scalable, simulated platform. The focus is on the development of verified quantum circuits for certain problems. Perfect qubits have no decoherence and no errors in the quantum gates. He is excited about quantum computing as the ultimate accelerator giving access to problem domains that are out of scope in conventional computing.   For any comment on this paper, you can contact him at koen.bertels@qbee.com.

     \item[Emma Turki] is an associate professor and Senior Researcher at Rouen Normandy University, in France. She is an expert on Physical Oceanography, Coastal dynamics and climate by developing innovative models and Earth Observations techniques of remote sensing. She has had collabotion with NASA and ESA and is considered one of the international experts on EO.

    \item[Aritra Sarkar] is doing a Post-Doc at QuTech, in the department of Quantum and Computer Engineering at the Delft University of Technology. He received his Bachelor of Technology (B.Tech), in Avionics, from the Indian Institute of Space Science and Technology, Thiruvananthapuram, India, in 2013. He worked in the Indian Space Research Organisation as a Scientist  till 2016. He completed his Master of Science (M.Sc), in Computer Engineering, from the Delft University of Technology, Delft, Netherlands, in 2018 and then joined the Faculty of Electrical Engineering, Mathematics and Computer Sciences, to obtain the title a Doctor of Philosophy (Ph.D.) in quantum algorithms. His research interests lie in quantum acceleration for experimental algorithmic information theory applications in artificial general intelligence and genomics.
    
    \item[Imran Ashraf] received his Ph.D.~in Computer Engineering from Delft University of Technology in 2016. The focus of his research was advanced profiling, accelerator-based computing, communication driven mapping of applications on multicore platforms. In 2016, Imran started working as Post-Doctoral Researcher on Intel funded project at Quantum and Computer Engineering department, QuTech, TU Delft. His research focused on simulation and compilation techniques for quantum computing and scalable architectures for quantum computers. Currently, Imran is working as an Associate Professor at the Computer Engineering Department, HITEC University, Taxila, Pakistan.
     \item [Tamara Sarac] received her Ph.D.~in Engineering Science from the University of Louvain-la-Neuve in Belgium. Up to date she is involved in material science with positions in research and development sector and over the last years she is oriented toward an application of quantum computing in chemistry. She is responsible for all the quantum chemistry algorithms and analyses that needs to be done in the context of a joint collaboration with the other authors.
\end{description}

\end{document}

%% file: 01-Introduction.tex
\section{Introduction}

Quantum computing as a scientific goal was launched shortly after a talk that Richard Feynman gave in 1986 to highlight the advantage of simulating quantum dynamics on controllable quantum systems. 
This inspired the quantum hardware community to look at these challenges and they started manufacturing the quantum bit, referred to as the qubit.  Given the exploration of a completely new kind of scientific and computing-oriented line of research, one of the first assumptions they made was to assume that the theoretical behaviour of qubits can be defined as coming from \textbf{perfect qubits}. Those concepts allowed the physicists and mathematicians to focus on the expected behaviour and assumed the qubit did exactly what they wanted. We will reintroduce this concept in the updated long-term road map.
\begin{figure}[bht!]
    \centering
    \includegraphics[width=0.8\linewidth]{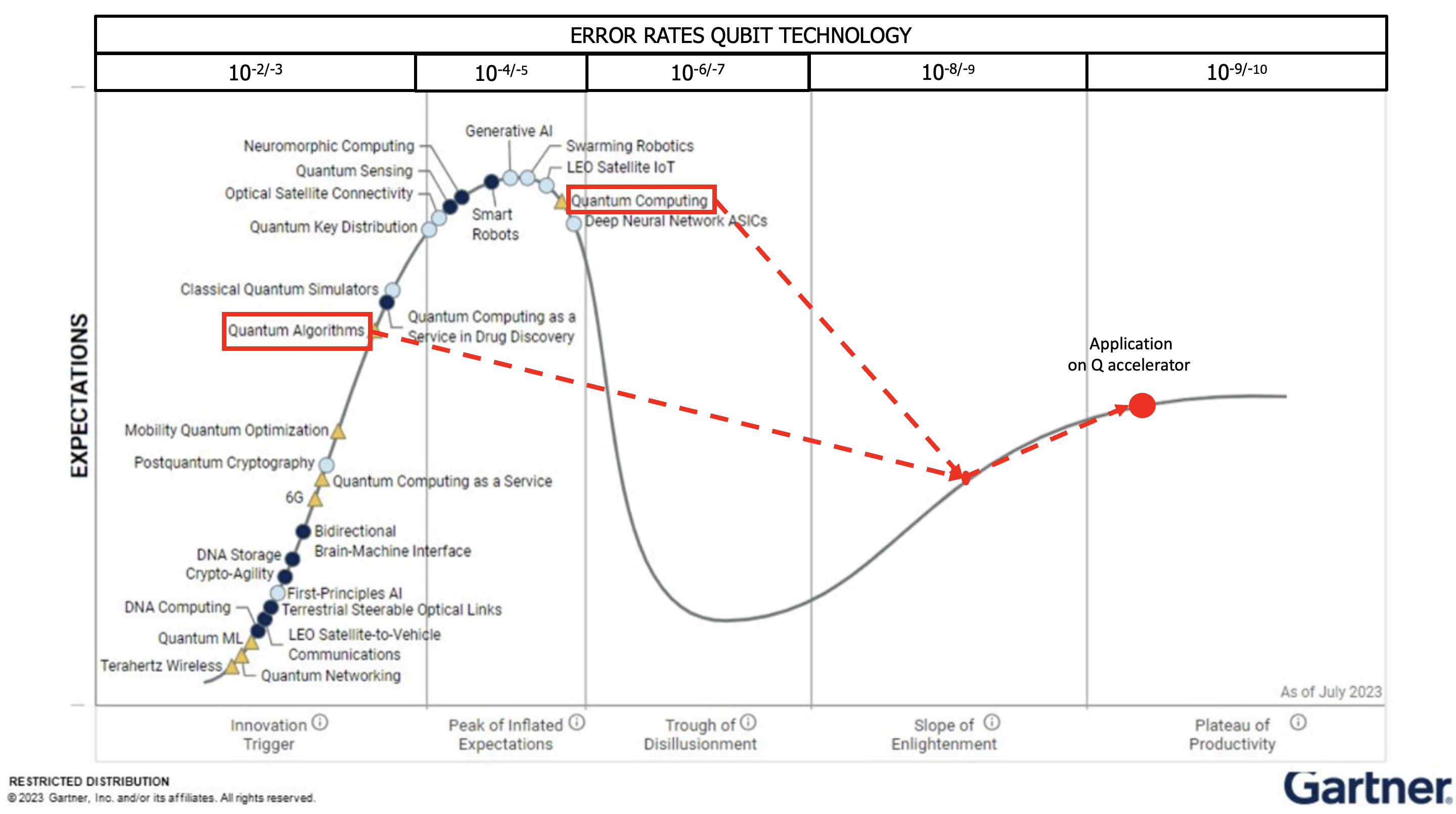}
    \caption{Gartner's Quantum Computing Hype cycle}
    \label{gartner}
\end{figure}
The theoretical and applied benefits of the first quantum computing algorithms, formulated in the 1990s, established this field as a concrete research direction.
In addition, quantum computing got a huge boost from the slowdown of  Moore's law of transistor scaling.  We are now at a \textbf{2nm-scale of CMOS-transistors}; it is quite uncertain that we can go even smaller, and still get more compute power.  This pushed major industrial players to substantially invest in the development of a quantum computer. The main ideas of our quantum research came out of the collaboration between Intel and the Quantum Computer Architecture (QCA) research group at  TU Delft. 
Before we briefly explain part of the full stack, we want to highlight an important observation that the American research and consulting company, Gartner, has made while following several technologies, and this for several decades. There is certainly repetitive behaviour in  the process of any new technology in the market.  Figure \ref{gartner} was found summer 2023 but it is no longer on the Gartner website.  Given the existing need to a more powerful computer technology, we use it as the basis for our reasoning. Grover defined the the Hype cycle in the following way, and we apply it on quantum computing potential. First and due to \textbf{triggering the desire for innovation}, an exaggerated enthusiasm with potential users is created. Error rates are around $10^{-2/-3}$.  This  often leads to  an over-investment in the early-stage of any technology. The next step is that \textbf{the peak of inflated expectations} stops because the technology has not yet sufficiently matured and fails to deliver the promised results. Error rates have a bit improved to $10^{-4/-5}$. This failed peak is then followed by a disinvestment in the technology. This step is called \textbf{the trough of disillusionment}. The error rates have hopefully improved reaching $10^{-6/-7}$. After some time, companies and universities pick up the core ideas and continue their development. This step is called \textbf{the slope of enlightenment}.  The error rates keep on improving and can be around $10^{-8/-9}$. In many cases, that particular step leads to a marketable product, called \textbf{the plateau of productivity}. The error rates should then have achieved $10^{-9/-10}$ or higher.  \footnote{https://www.gartner.com/en/research/methodologies/gartner-hype-cycle}  None of the qubit technologies delivered  the long anticipated results, leading to the hype cycle peak. Hence leading to the disinvestment phase for quantum hardware. This is also due to very large investment needs and the uncertain results for many qubit technologies.  Some of the authors of this paper have worked, as computer engineers, on the quantum hardware part and they are convinced that it will take another 10-15 years before any qubit technology matures and can be put in the market. The main conclusion we are formulating is that one has to be careful when assessing the current scientific results  and not be discouraged by inflated quantum expectations.\cite{mckinsey2021}\cite{mckinsey2022}  In this context, Gartner mentions on their website  that CIOs should only expect results in at least 5 years from now.\footnote{https://www.gartner.com/smarterwithgartner/the-cios-guide-to-quantum-computing} We personally think it will take a bit longer as the core technology is not yet working as it should and there are still many unknowns in terms of quantum algorithms.  That is also the reason why an increased rise for quantum software. We have to stay careful as also the quantum software branch is also  part of the hype cycle.   Universities and companies have to be careful about their investments in terms of people and knowledge. 

In this paper, we first present the overall quantum hardware challenges, irrespective the kind of qubit technology being used. We also present an updated road map for the next 10-15 years, based on the \textbf{Perfect (qubits) Intermediate Scale Quantum} approach as first described in \cite{9537173}. We the present three examples of quantum algorithms that are being developed by the authors of this paper, namely Earth Observation, quantum computational chemistry and quantum genetics.  We conclude the paper by a refined, and fully updated path forward not only for the QC research community but basically for all scientists that need better performing computers for their fields. That is why we present three examples for which we are developing a quantum computing logic, called \textbf{QC-logic}, version.

%% file: 02-FullStack.tex
\section{The Full Stack}

In Figure \ref{fullstack}, the full stack~\cite{9116502,bertels2021quantum} is shown which  describes the most important layers that are now needed in any quantum device. As stated before, we have started working on these layers more than 10 years ago and it is now one of the standard architectural frameworks. It describes the minimum required components for a complex quantum information system and the layering enable for a separation of the concerns. However, we believe that  many breakthroughs are still needed, which require a lot of collaborations including between research and industry.
The architectural framework can be summarised as follows:
\begin{itemize}
    \item \textbf{Quantum Applications and Library -} At the highest level, the application is formulated into a set of quantum algorithms, using the quantum logic primitives from the quantum library. The algorithms describe how many qubits need to be used and what operations should to be performed among them.  We have started working on quantum genetics, quantum computational chemistry and quantum earth observation (EO).
    \item \textbf{Quantum Verification \& Validation -} We also need many more tools that will allow quantum application developers to verify and validate the results of the quantum circuit. Comparison against classically computed results stays very important in the early phase of quantum software development.
    \item \textbf{Quantum Programming Languages -} The application is often expressed in a high-level programming language that many companies and universities have developed, such as OpenQASM~\cite{ibmqasm20} and  OpenQL language~\cite{khammassi2020openql}. The OpenQL back-end compiler generates cQASM~\cite{khammassi2018cqasm}, a quantum assembly language.  However, there are several languages and compiler platforms available, developed in the USA, Europe and China. Each quantum language will always generate a Quantum Assembler version, which we call QASM. One of them is the public domain platform Qiskit, initially developed by IBM, and used by a large number of researchers world-wide, feeding the results to a quantum library.
\begin{figure*}[bht!]
\centering
    \includegraphics[width=0.8\linewidth]{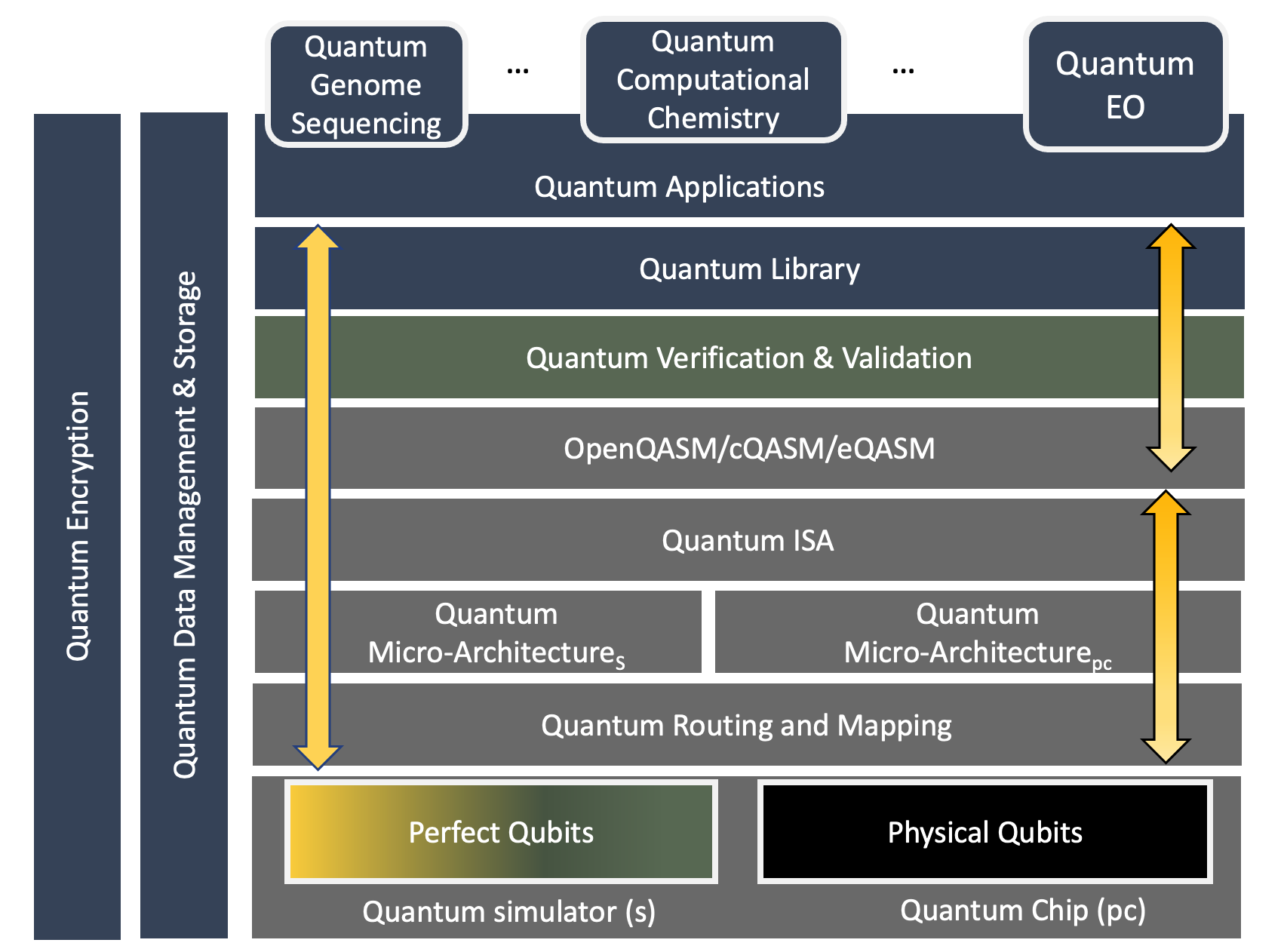}
    \caption{Full Stack and Qubit technologies}
    \label{fullstack}
\end{figure*} 
    \item \textbf{Quantum ISA -} Currently, quantum chip developers use around 20-30 quantum gates to express any quantum operation on the superposed or entangled qubits. Much more application development is needed to identify, test and implement new quantum gates. It is a joint decision to implement the new gate as a hardware extension or to have the compiler translate the new gate as any combination of the universal quantum gate set.
    \item \textbf{Quantum Micro-Architecture -} A micro-architecture receives the QASM-instructions and does an internal processing before sending it to the lowest level which can be a hardware quantum chip or a quantum simulator.  The non-exhaustive list of competing technologies is depicted in this figure as well, for instance, for superconducting qubits~\cite{fu2017experimental}. Important to notice is that the micro-architecture for the perfect qubit simulator simpler is than the one for any of the hardware qubit chips, hence the 's' or 'pc' indicator in the figure.
    \item \textbf{Quantum Routing and Mapping -} Figure  \ref{fullstack} has changed a little bit as quantum routing and mapping of qubits has been added.  Depending on the qubit technology, qubits have to be placed close to each other for n-qubit gate operations. Depending on the kind of hardware technology, the routing and mapping of qubits will be completely different.
    \item \textbf{Quantum Chip -} At the lowest level, we can either use a hardware quantum chip or a simulator. Given the current quality of qubit chips, we have developed our own simulator that can execute the QASM-instructions.~\cite{budhrani2020quantumsim} On a normal computer, we can execute up to 28 superposed qubits and are around 70 or 80 qubits when using very large memories of more powerful servers.
    \item \textbf{Quantum Encryption and Quantum Data Management -} Two aspects are very important for any layer, and they are encryption and quantum data management.  It is not clear yet how they will be implemented or interacting with any of the layers, but that is the main reason why they are represented in our full stack figure.
\end{itemize}

The full stack of a quantum computer or quantum accelerator is world-wide being followed and still a lot of collaborative work is needed to achieve sufficiently large compute platforms on which we can execute real-world applications.

%% file: 02-Problems.tex
\section{Quantum Hardware Challenges}\label{QChallenges}

In this section, we review some of the more important open issues  from the perspective of quantum hardware.  
Our main starting point is a talk, John Preskill gave in 2018, about the challenges in quantum computing. In two  accompanying papers~\cite{preskill2018} \cite{preskill2021quantum}, Preskill targets both  active researchers in quantum computing as well as  a broader audience. 

\begin{itemize}
     \item \textbf{Diversity in quantum technology:} Many technologies are still competing with each other to make the best qubit. It is uncertain what technology will win. It is to be expected that many other qubit technologies will also disappear in the coming years. We just need to remember that in transistor development, it took the entire world around 40 to 50 years to reach a VLSI-level of transistor production, which is based on an idea formulated in 1936. Given the current quantum market, the qubit technologies that seem to be performing well are Ion Traps, Photonic qubits and Superconducting qubits.
    
    \item \textbf{Number of qubits:} 50 hardware qubits is what Preskill calls a significant  milestone in the sense  that the quantum hardware community is capable of achieving a level of performance beyond the capacity of classical computers.  The number 50 is motivated by the maximum number of qubits we can simulate using supercomputers. The main memory, in supercomputers, stores all the information of the full quantum circuit, implying the basis states and the amplitudes that define qubits. However, with all the problems qubits still have, it is very doubtful how to interpret and assess the computed results.  Preskill is therefore suggesting to consider up to 100 hardware-based qubits  to substantially improve their quality, such that a choice can be made regarding the best qubit technology among many competing ones, and to continue developing and improving that choice.
    
    \item \textbf{Coherence of qubits:}  The  coherence time is the length if time the information contained by the qubit is accessible and usable. This coherence time varies significantly depending on the quantum technology used. For different quantum technologies, we have found values that go from a couple of seconds to multiple minutes.  What is important to realise is that the coherence time needs to be substantially longer than the time it takes to execute the full quantum circuit because otherwise all intermediate or final results will be lost.
   
    \item \textbf{Quality and number of quantum gates:} The accuracy of the quantum gates is also a major problem as the error rates are way too high to implement any meaningful application that can be tested on its quantum formulation.  
    Preskill suggests to limit it to 1000 gates as the noise will be so high that it is difficult to assess the quantum results obtained. This is certainly meaningful for qubit development but we need to look at many other aspects too to have substantial improvements for any scientific, technological and in general society-relevant applications. 
    
    \item \textbf{Quantum error correction:} Given  the errors in the computations and the overall behaviour of any qubit technology, there is a need to correct the quantum (intermediate) results such that the qubit states do not incorrectly accumulate all the errors of the preceding computations.  These days, the error rates are $10^{-2}$ or $10^{-3}$ and it is interesting to understand what the qubit engineering researchers want to achieve in, for instance, 5 years from now. In CMOS, we are used to having $10^{-15}$ and that is far too ambitious for the next 10 years. But important to know is when can one expect to reach $10^{-7/-8}$, which most likely is still 10 years or longer away.  We should not forget that, when compared to classical hardware, error rates of $10^{-15/-16}$ are normal and can be handled by the classical hardware.  
    
    \item \textbf{Logical qubits:} An important attempt was to formulate logical qubits based on multiple hardware-based qubits. The goal is to have an overall qubit behaviour which is more stable and scalable. One approach was  based on Surface Codes for which we need 49 hardware qubits to have one logical qubit.\cite{LLao2019} \cite{GoogleQAI} So, also Surface Code and other logical-qubit approaches will have to be postponed or substantially reduced in size.

\begin{figure*}[bht!]
    \centering
    \includegraphics[width=0.9\linewidth]{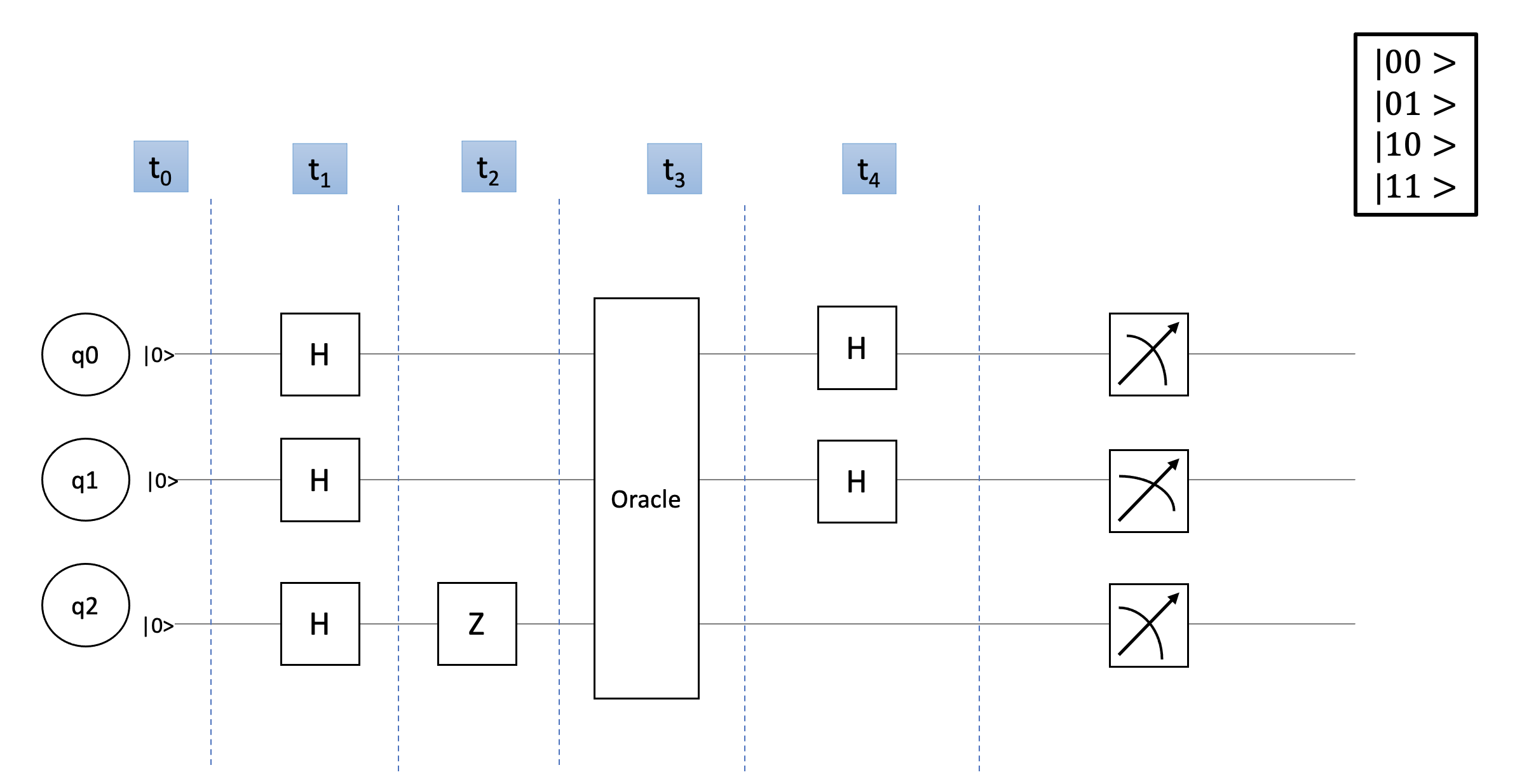}
    \caption{Bernstein-Vazirani (BV) algorithm,}
    \label{bernstein}
\end{figure*}

    \item \textbf{Quantum oracles:} While oracles are not specifically a quantum hardware problem, it is nevertheless a crucial component.  To apply and assess  any quantum algorithm in practice, it is needed to holistically consider the oracle as well, the part considered as a black-box in the original formulation. When all of the details of a quantum circuit or algorithm is not known, or when certain gates depend on the specific initial value of the qubits, the oracle as a useful technique. A famous example is the Bernstein-Vazirani algorithm, which is shown in Figure~\ref{bernstein}, where the content of the oracle in the circuit changes in view of the initial state of the qubits. For instance, for the initial qubit state \(\ket{Q_0Q_1}=\ket{00}\), the black box turns in a CNOT-gate  where  \(\ket{Q_1}\) is the control qubit and \(\ket{Q_2}\) is the target qubit, leading to a final result of \(\ket{Q_0Q_1}=\ket{01}\) for this particular initial state.

    \item \textbf{Variational heuristics:} More recently a lot of focus has been on variational heuristics where a parametric circuit is trained on a quantum computer using a classical optimiser.  Lack of universality of quantum software constraints the evolution of important quantum algorithms such as Variational Quantum Eigensolver (VQE) and Quantum Approximate Optimisation Algorithm (QAOA). We need substantial research being done in that direction to reach wide adoption of these tools.
    
    \item \textbf{Scalable fabrication of qubits:} It is not yet conceivable to develop fabrication technologies as long as there is no understanding and agreement on what quantum technology can be used to produce good qubits.  It is unlikely that all quantum technologies will survive and it will most likely be the role of a small number of big players that will outline what technology will reach the market. 
\end{itemize}

The NISQ-approach is clearly the most realistic approach and more likely a very promising direction in which quantum hardware can continue researching the development of qubits. The list of challenges that we discussed in this section clearly focuses on a lot of engineering aspects that can be solved by researchers with a quantum hardware background.
There is also no agreement on how we define the words \textbf{good} and \textbf{scalable} where scalability is needed to compensate for the errors  in qubit behaviour as well as other required elements such as dilution refrigerators and qubit connectivity. We believe that, 50 really good qubits with error rates of $10^{-8}$ is the appropriate threshold for achieving descent computational results. However, as we are now only capable of making 50 qubits with error rates of $10^{-2}$, it is very problematic for the world. 

In the context of all these issues, it is very interesting to hear about the 127-qubit chip that IBM recently announced.\footnote{https://research.ibm.com/blog/127-qubit-quantum-processor-eagle}  We are evidently very interested in finding out what the error rates are of this chip and we can only hope that we are substantially moving away from the $10^{-2/-3}$.  
That is why we propose an alternative approach where a wider interest community can start looking at the development of quantum solutions and algorithms.

%% file: 03-00-QRoadMap.tex
\section{The Quantum Roadmap}

In the beginning of quantum computing, many different hardware approaches were discovered to make quantum bits, called qubits. As we write this paper, there are still several competing technologies to make a good quality qubit.   The long-term goal is to fabricate a quantum chip with a high number of good-quality logical qubits, which can be implemented with our without quantum error correction (QEC).  We will focus on \textbf{FTQC} -  Fault-Tolerant Quantum Computing, \textbf{NISQ} -  Noisy Intermediate-Scale Quantum computing, \textbf{PISQ} -  Perfect (qubits) Intermediate Scale Quantum and  \textbf{NFTQC} - Non Fault-Tolerant Quantum Computing.

    \begin{figure*}[bht!]
    \centering
    \includegraphics[width=0.9\linewidth]{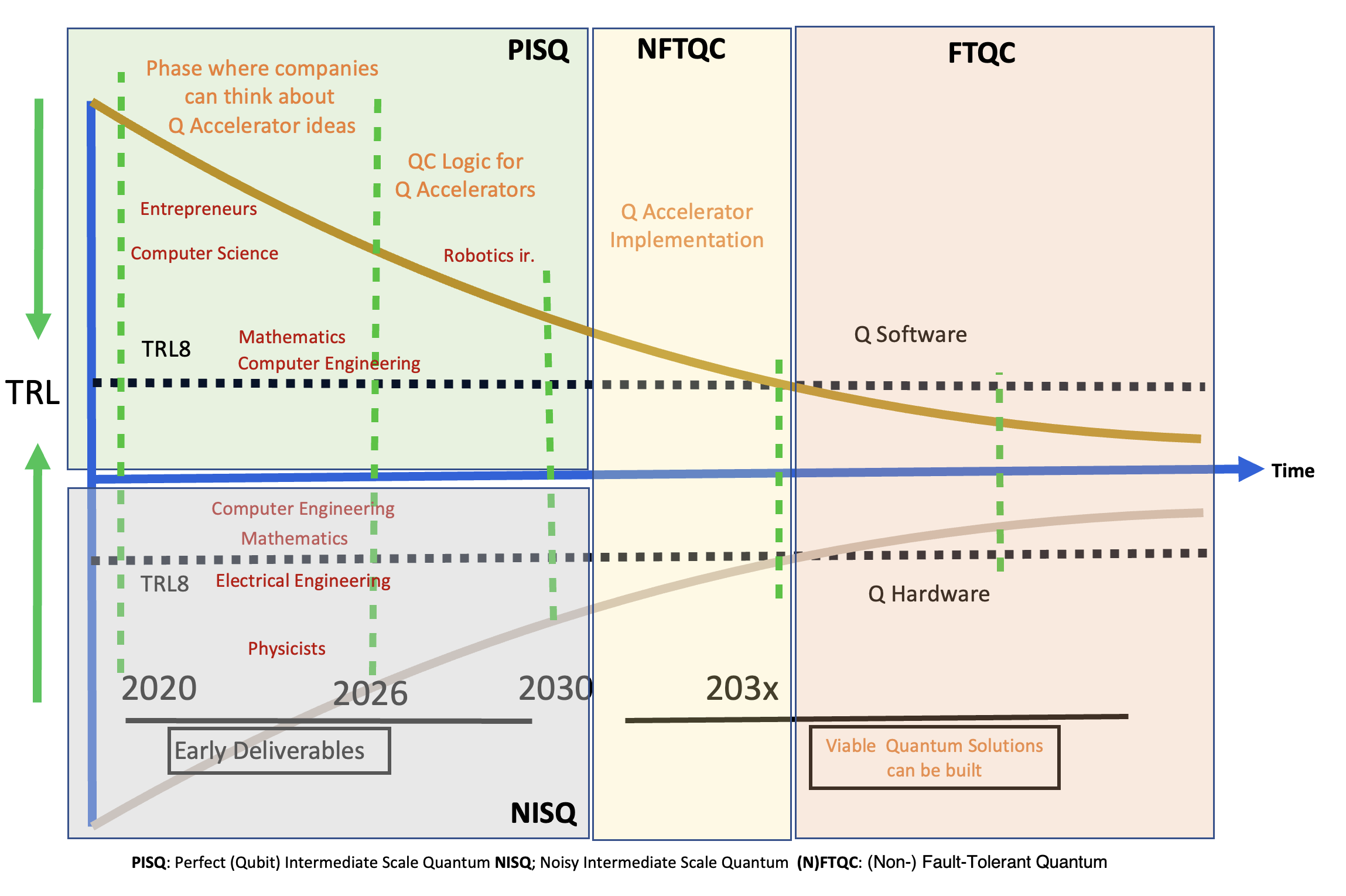}
    \caption{Long term inter-disciplinary collaboration}
    \label{LTQC}
\end{figure*} 

We refer again to an updated Figure \ref{LTQC} that shows the long-term view of how different research fields could  collaborate.   From a computer engineering point of view, we are expecting error rates up to  $10^{-5/-6}$, and, currently, we are not doing much better than $10^{-2/-3}$. Also, coherence times vary from very short times up to several minutes. As technology keeps developing, we need to reach much better error rates and longer coherence times. It is always difficult to predict how fast a technology will evolve to good and useful quality.  As shown in Figure \ref{LTQC}, we describe a 10-15 years horizon, where NISQ and PISQ represent complementary approaches, and we move to FT quantum computing in later phases. We briefly describe the 4 distinct steps. And just to be clear, we do not believe in very fast progress, it takes time to move forward, but quantum technology has a promising future.

\begin{itemize}
     
    \item \textbf{NISQ - Noisy Intermediate-Scale Quantum } - In 2018, the FTQC-approach was replaced by  NISQ, a term formulated by J. Preskill. In general, the word \textbf{`noisy'}  refers to 2 complimentary challenges, namely the quantum gate errors and the qubit decoherence. Errors refer to incorrect gate execution on the analogue phenomena of any qubit and decoherence refers to the fact of going back to the ground state of a qubit, thereby loosing all obtained information in the qubit.
    The term \textbf{ `intermediate-scale'} refers to the number of hardware qubits, such that we cannot exhaustively simulate them on a classical computer.   For various reasons it appears that NISQ did not deliver as much quantum advantage as expected. This is being discussed within the quantum community. In this paper, we introduce an alternate but closely related concept, called \textbf{PISQ} which stands for \textbf{Perfect Intermediate Scale Quantum } computing as will be explained later.    Using the PISQ-methodology,  developers, when testing their quantum circuit, can use up to 70 qubits in superposition on a classical supercomputer.  There are multiple results from the PISQ-approach. The first, and very important for industry, is the realisation of quantum circuits for their organisation. The second, and more system-oriented results, are the definition of new quantum gates and the testing of new quantum micro-architecture, implemented using classical technology.
    The goal of NISQ was to improve the competing qubit technologies such that some maturity is achieved. It is very difficult to find a fundamental solution to one of the biggest quantum challenges, which is coherence and much less errors in the quantum operations.  When one of the authors of this paper visited Prof. A. Morello from UNSW, Australia, Morello already expressed himself like that in February of 2023. 
    What is interesting, is the fact that certain research groups focus on a small number of qubits but try to perfect them as much as possible. So coherence times of multiple minutes and quantum circuit gates executed and much better results are achieved. It is our opinion that certain companies such as INTEL, IBM, and others need to focus on the number of qubits but they should also pay attention to many university-based QC-groups and combine those results with the large number of qubits.

    \item \textbf{PISQ - Perfect (qubits) Intermediate Scale Quantum} - We introduced this PISQ-label to improve the development of end-user quantum applications. The goal is to focus much more on application development for which we need very large and powerful computer platforms. PISQ  is independent of any of the qubit technologies but looks exclusively at the functional outcome of any quantum circuit. One abstracts away all the decoherence and quantum errors that any hardware qubits approach still has. It allows us to focus on quantum computational logic. PISQ is a very user-friendly way to develop quantum versions of the functional software anyone will need, once the quantum computer or accelerator will reach the market.  Many more simulators are being developed to execute PISQ-based algorithms, even though few people will call it PISQ. It is an opening to any industrial or societal-relevant topic for which no quantum algorithms or versions have been developed. Multiple benefits can be found when approaching the PISQ-methodology.  The first is that one focuses on functionality of the quantum circuit one needs for any application, such as genetics and chemistry.  The second benefit is that new quantum gates can be defined and used. It is absolutely vital that we increase the number of quantum gates above the 20-30 that are part of the universal quantum gate set. This is also important for the micro-architecture and the hardware implication of any qubit chip.
    
    \item \textbf{NFTQC - Non-Fault-Tolerant Quantum Computing}: This is an approach which can follow any NISQ-based result and can be based on any PISQ-algorithm developed.  Given the state of the art of any qubit technology, the NFTQC-community can further perfect and expand the quantum circuits defined by the PISQ-methodology and assess how fault-tolerance modern quantum hardware is, hence the direct link with the FTQC-community.  Any quantum algorithm can be executed on, for instance, a PISQ-simulator such as from IBM. It also allows to assess how good the hardware qubits are when exactly the same circuit is executed on a hardware quantum chip. This double check allows us to give good feedback to the Q physicists working on any qubit-technology, but it also will allow any other researcher to identify the correct and useful new quantum gates needed for their specific scientific needs. An important side effect is that it should lead to a more efficient micro-architecture as there will always be a connection between classical hardware and any quantum accelerator that will be developed.
   
    \item \textbf{FTQC - Fault-Tolerant Quantum Computing} - This abbreviation stands for Fault-Tolerant Quantum Computing. This is still very focused on the quantum hardware challenges presented earlier in the paper.  One of the important approaches was the creation of a logical qubit. Surface Code or Color Code were, for a  while, the most promising options to make a logical qubit. This is still a very dominant approach but it no longer dominates the quantum hardware community. The most important reason is that one needs e.g. 49 hardware qubits to make one logical qubit.\cite{GoogleQAI} So, there is high cost in terms of number of qubits.  The scalability is realised by implementing surface-code, color-code or other approaches. Now, many references can be found looking at Error Mitigation but it is not clear how easy of difficult they are, nor how scalable they will be.  Many quantum chips offered these days have a relatively low number of hardware qubits. So, scalability is not yet achieved by anyone. But maturity will be achieved, as an outcome of the NFTQC approach, but it is not clear yet how many qubits we will need for what application. An important interaction is with researchers from the NFTQC-community. Through their algorithm development by the NFTQC-researchers, new gates and quantum methods are developed and formulated. This will allow the FTQC-researchers to include, for instance, a hardware implementation for any quantum chip that is being developed.

\end{itemize}

One of the conclusions we formulate for this section is that we need to encourage many more researchers, working for universities or industrial companies, to work on QC-logic based quantum algorithms. The expected benefits will be new quantum gates, new quantum methods, improved quantum micro-architecture, quantum programming languages, quantum operating systems, quantum databases, etc.  In the following section, we introduce some application fields on which we have already been working. It is now also extended with Earth Observation, one of the topics that the European Space Agency, ESA, is interested in.  Our work can be interpreted in the context of both Earth and Mars exploration, where EO and chemistry will be both applied.


%% file: 04-00-QExamples.tex
\section{Quantum field examples}
A recent Wikipage presents around 20+ articles that are always referred to.\footnote{https://en.wikipedia.org/w/index.php?title=Quantum_algorithm\&oldid=1189520247} It clearly shows there is a great need for many more researchers, from whatever field, to enter the QC-logic initiative, for which a specific mathematical approach is needed, namely  Tensor mathematics.\cite{desurvire2009}\cite{Tensor}  Given the full stack as shown in Figure  \ref{fullstack}, any scientific field should open up to quantum-driven scientific research, going beyond the qubit technology aspect.  

We now briefly introduce the different fields we are focusing on, and that scope can be made much bigger, assuming that universities, companies and others join this approach.  We should not forget that there is no alternative computer technology begin developed that, potentially, has the same compute power as a quantum computer or accelerator. We also have to be aware that it will take around 10-15 years before an established, and scaled approach is available to develop many more quantum algorithms. The PISQ-methodology is just a first approach such that any researcher can focus on the functionality of the quantum circuit, abstracting away all the challenges that still exist at the quantum-hardware level.


\subfile{04-01-EO}
\subfile{04-02-Chemistry}

\subfile{04-03-Genetics}

%% file: 04-01-EO.tex
\subsection{Earth Observation}

Earth observation (EO) could be defined as the collection and presentation of all satellite-image-based information related to our planet, focusing on natural systems as well as human-driven actions from locations all around the world. The goal is to have a better understand for their spatial and temporal evolution. Earth observations gather the local ground level, acquired from airborne, and the global level, acquired from remote sensing platforms as satellites. The remote sensing approach has gendered the development of powerful scientific tools. These tools provide data on greenhouse gas emissions, natural systems such as forests, sea levels, and glaciers, and they also look at all aspects of urban environments.

EO data plays a crucial role in monitoring  the planet, tracking the spatio-temporal evolution and responding to climate-related disasters, such as flooding, wildfires, and storms. The goal is to save lives and reduce property damage.  The huge amounts of data, provided by satellites in space, are around 150 TB per day. These images provide information in the time and spectral domains. Those images are manually or automatically  processed, and utilised by users who translate that information in operational processes. This extraction is very challenging and requires specialised skills and resources.  In classical technology, deep learning-based methods gained prominence in analysing images for real-time applications, which has been investigated recently in several works such as convolution neural networks (CNNs) proved to be efficient for various image processing and scene classification tasks. 

In \cite{Reichstein2019DeepLA}, the authors address these challenges by developing integrative hardware modelling  with machine learning (ML).   The performance and accurate use of deep learning (DL) models for the classification of image scenes is described in \cite{MA2019166}, taking the real-world scenarios, land-use, the detection of objects, and semantic segmentation into account. The authors also highlighted the challenges in training the supervised DL models such as CNN. 
Then, the use of trainable multi-layer networks has been proven to be successful to extract features from an image of remote sensing image scene classification. Indeed, the deployment of convolution, pooling and connected layers are useful for improve the image extraction (\cite{ZOU2015}). \cite{ZHANG2019}  have developed fine tuning of CNN models and they have highlighted their efficiency for the scene classification task. 
 
 However, the implementation of such methods requires a high-performance computational facility as many parameters are used in training such models. For example, the training of deep neural networks (DNNs) with more layers is a complex task in terms of optimisation and tuning of parameters . As the number of parameters increases, the convergence of the optimisation process is difficult during training the model. The deeper neural network architectures with a large number of parameters often lead to multiple local minima. Computationally expensive hardware and memory-intensive methods are required to manage a large number of parameters in a deep model with more layers . 

\textbf{Quantum approaches of EO - } This is where the technology provided by quantum accelerators becomes very useful. We have to be aware of the fact that huge amounts of data cannot be processed yet, whether we target a hardware platform or a perfect qubit simulator.  That is why scribbled in terms of number of qubits is one of the challenges of the quantum hardware.  We also need to be aware that any application will most likely be a combination of existing classical software with quantum extensions running on a quantum accelerator. What is relatively clear is that, for many fields,  large numbers of qubits are needed, which not a single university or company can already offer.  Such applications where multiple computing platforms are combined are called hybrid.  

Remote sensing images create about 150 TB of data per day. \cite{Sebastianelli2022} Remote sensing is defined as \textbf{"the acquisition of information about an object or phenomenon without making physical contact with the object, in contrast to in situ or on-site observation"}. \footnote{https://en.wikipedia.org/wiki/Remote_sensing}
The largest image processing on which a QC-logic algorithm is used, can be found in \cite{Chalumuri2021}. The authors describe a system consisting of 9 qubits on which a quanvolutional (or quantum convolutional) neural network is running. The image contains 512 x 512 pixels. The main advantage is clearly driven by the huge amounts of data that need to be processed.


Important in this context is the work of F. Leymann who is building a compiler that can target any qubit-technology. He explicitly mentions the pre- and post-processing steps, that are usually forgotten when compiler an executable version.  Those two steps are a heavy workload for any compiler, targeting any qubit technology.\footnote{https://scholar.google.com/citations?user=qRnegTsAAAAJ}  The different steps in any SIC algorithm consist of the following:

\begin{enumerate}
    \item \textbf{Selection of training examples:} An important part of the entire SIC logic is the selection of a representative set of training data. We will be using neural networks in a supervised way so it is very important to have good training data and a relevant set of testing data.
    \item \textbf{Data pre-processing:} An essential pre-processing step is to generate a stable and scalable version of the data. There exist many public databases as can be found in \cite{7822014} and many databases belong to private companies. Many satellite images have different kinds of data items. We need to be able to process and interpret the different layers in an image. The range of data goes from a very high level to maybe even at the \textbf{ meter} level. 
    \item \textbf{Feature extraction:} An important next step is to have a  numerical model of the data which is being processed while preserving the original but raw data set. This can be done in a manual or automatic way. \cite{Sublime2019} 
    \item \textbf{Classification:} Using neural networks or any other way of classifying the data, there are two main  approaches. It can be done in a supervised or unsupervised way. Supervised classification can be done through the use of maximum likelihood, artificial neural networks, minimum distance, and others.  The unsupervised classification tries to group pixels into unlabelled classes, after which human intervention will put meaningful labels on these groups. Examples are K-means and auto-encoders, restricted Boltzmann Machines, etc.
    \item \textbf{Post-classification processing:} Expert rules and human interpretation is often needed to improve the quality of the classifications. 
    \item \textbf{Evaluation of classification performance:} There are multiple performance metrics such as reproducible, robustness, full content use of the data, uniform applicability, and objectiveness.  None of the algorithms so far can comply with all these metrics.

\end{enumerate}

 Now that we have  clear idea of the different steps in any EO-algorithm, it is now time to briefly discuss some quantum tools that have been developed by other researchers.  The first thing to do is to define the metrics that we will be using to assess how good the results are and what challenges need to be solved before we can have a real progress in the QC-logic for Earth Observation. It should not be astonishing that very few papers report on all these measures.  Another important feature is the implicit parallelisation that quantum devices will do for any quantum circuit. It is very difficult to assess this, using a low number of qubits. So, this parallelisation will come back in a later, more mature phase of qubit technology.  When using the full stack as the main reference structure, at least the following metrics  need to be defined and used  for any EO-algorithm we will develop.

\begin{itemize}

    \item \textbf{Spectral resolution of satellite images}: For EO, we will be working on satellite images, but the generated amounts of images is around 170 TB per day. So, it does not make sense to look at those amounts. Resolution is an important feature of the images as it is also defined by the spectral aspects of the images.
    
    \item \textbf{Accuracy of classical and quantum EO-algorithms}: Classical algorithms can be defined in terms of accuracy in terms of feature detection. We can then compare the results obtained fro a quantum tool with those numbers. Those numbers can vary from 75\% to 95 \% for the quantum algorithms, compared to the classical ones.
    
\end{itemize}

%% file: 04-02-Chemistry.tex
\subsection{Quantum Chemistry}

Chemistry is expected to be among the first fields to benefit from quantum computer development. 
Markets  are always pushing for producers to design new chemicals or materials fulfilling desired performances and properties, processed and produced in a cost effective, as well as ecologically accepted manner. To be able to do so, many physicochemical phenomena need to be understood and  issues in predicting matter and compound behaviour need to be tackled. In general, computation is a very powerful tool in chemistry. Simulations are used to replace or complement experimental work, which decreases cost and increases efficiency and even accuracy.  

Even the strongest classical computers in use nowadays are limited in precision when used to describe molecular structures and chemical reactions. A good example is the so called \textbf{the electronic structure problem}, describing the issue of calculating the ground-state properties. For instance,  the lowest energy state properties of a chemical system. Electrons arrange in atomic and molecular orbitals in a specific manner following the laws of quantum mechanics and electron configuration of a molecule determines many physicochemical properties.\footnote{The atomic/ molecular orbital is a mathematical function of the most likely position of electron in an atom/ molecule.} Quantum mechanics is in the core of chemical systems and processes but exact application of the laws of quantum mechanics to these systems is limited by its computational complexity. Schrodinger equation is used to describe wave functions of a quantum system. However, the complexity of computing the solution of a Schrodinger equation grows exponentially with the number of electrons in the system. Quantum computers come in a natural as possible way to the solution of this problem. It was  originally stated by Feynman who first defined that we could use quantum computers built on quantum hardware to simulate and hopefully better understand the properties of  quantum systems.  
Development of quantum computers is therefore directly related to the world of chemistry, starting at solving computational issues that are out of the range for classical computers. Quantum computers are expected to be able to simulate electronic structures of more complex chemical systems but also to do it in a more accurate way. 

Quantum computers can be seen as a new tool in computational chemistry and implementation of it is challenging on its own. Number of things can be addressed in reaching the final goal - obtaining advantage of quantum computer usage in development of new materials, drugs, catalysts or understanding chemical processes. The procedure for calculation of a desired parameter usually starts with mathematical representation of a system of interest, usually using an orbital basis set (this is a set of calculation-friendly mathematical functions, linear combination of which approximates orbitals ). One needs to instruct a computer how to perform a computation by designing algorithms. In computational chemistry it implies choosing a method that involves approximation developed by theoretical scientist, in order to compute desired property for a given system. There is always a drive for applying methods that include less approximation, because they are more realistic and accurate, together with usage of extensive basis set (composed of increased number of basis set functions), because it describes orbitals with more credibility. Combining the two directs toward more exact solutions, but it is extremely computationally expensive even for simple systems. Quantum computer usage, implies information representation in terms of qubits and quantum gates, so mathematical representation of a chemical system and calculation method need to be encoded. Quantum circuit needs to be designed considering available resources, and implementation on hardware or simulator. For example, nowadays the mostly emploed algorithm is Variational Quantum Eigensolver (VQE), a hybrid quantum-classical algorithm. 

The previous is roughly listing challenges, potential objectives of a research calling for a multidisciplinary approach when aiming in obtaining the most benefits of quantum computers in chemistry. Industrially relevant applications can depend on progress on computation procedure in therms of developing new computational process or adapting the existing one toward the new tool (from classical logic toward quantum). Theoretical chemists, physicist, mathematician could contribute in quantum-computer friendly methodology development at a very fundamental level , while a computational chemist can think of the best way to correlate the new tool to the application needs, trough a most suitable encoding method or algorithm development. Experts from application domain can be supported by computer engineers or programmers, to come up with concepts that are suitable for implementation with respect to a qubit technology or simulator possibilities. Therefore, authors of this paper want to call for multidisciplinary approach in developing a quantum logic with respect to the application domain.

Many hardware issues are yet to be solved to enable advanced quantum chemistry (and many other) simulations, and these are mainly related to the quantum computer's system stability and error corrections. But the lack of the hardware capability should not stop the researchers from already developing theory and methods behind quantum chemistry algorithms and proofs of concepts tackling subjects and applications of interest.
In that sense, it is important to do so following  herein the described PISQ approach.

%% file: 04-03-Genetics.tex
\subsection{Quantum Genomics}

Understanding the genome of an organism reveals insights~\cite{sarkar2021} with scientific and clinical significance like causes that drive cancer progression, intra-genomic processes influencing evolution, enhancing food quality and quantity from plants and animals.
Genomics data is projected to become the largest producer of big data within the decade~\cite{stephens2015big}, eclipsing all other sources of information generation, including astronomical as well as social data.
At the same time, genomics is expected to become an integral part of our daily life, providing insight and control over many of the processes taking place within our bodies and in our environment.
An exciting prospect is personalised medicine~\cite{hamburg2010path}, in which accurate diagnostics can identify patients who can benefit from precisely targeted therapies.

Despite the continual development of tools to process genomics data, current approaches are yet to meet the requirements for large-scale clinical genomics.
In this case, patient turnaround time, ease-of-use, robustness and running costs are critical.
As the cost of the whole-genome sequencing (WGS) continues to drop~\cite{wgscost2018}, more and more data is churned out creating a staggering computational demand.
Therefore, efficient and cost-effective computational solutions are necessary to allow society to benefit from the potential positive impact of genomics.

In our research, we developed efficient solutions based on the quantum computing paradigm to the high computational demands in the field of genomics, specifically for genome sequence reconstruction.
Nucleic acids like DNA and RNA carry the genetic instructions used in the growth, development, functioning and reproduction of biological organisms.
These very long thread-like polymers are made up of linear array of monomers that encode the genetic instructions.
The four constituent nucleic molecules, adenine (A), cytosine (C), guanine (G) and thymine (T) allow simplifying the representation of both DNA and RNA as a single string with four distinct symbols while processing it as digital data.
The length of genomes varies greatly among organisms, for example, the human genome is approximately $3.289\times 10^9$ bases long (DNA has a complementary strand of base pairs, but is is a duplication of information).
Owing to this length, it is not possible to obtain the entire sequence in a single readout from the sequencing machines.

In order to sequence the organism, multiple copies of the DNA/RNA are broken down into fragments as sequencing machines are not capable of reading the entire genome at once.
Then, these fragments are sequenced using modern sequencing technologies (such as Illumina), which produces reads of approximately 50 to 150 bases at a time, with some known error rate.
Then these short strings are stitched back together - a process called sequence reconstruction.
Genome sequence reconstruction is primarily performed using two techniques, namely  \textit{de novo} assembly of reads and  \textit{ab-initio} alignment of reads on reference. De novo assembly is applied while sequencing a new organism, where the short reads are stitched back together based on the overlap between each pair.
This is computationally very intensive and is impractical for classical high-performance computing except for small micro-organisms.
For organisms with longer genomes, for example, humans, the alignment method is preferred.
Once the DNA/RNA is constructed for a species, for example, via the Human Genome Project, this is used as a reference for further individuals of the same species.
The whole genome is reconstructed by aligning the short reads on the reference genome.
Thereafter, the variation from the reference genome can be inferred to understand specific traits or abnormalities in the individual.

Since the principles of quantum computation are fundamentally different, we investigate the most basic algorithmic primitive for which we constructed the quantum kernel for acceleration.
We designed a new algorithm, quantum indexed bidirectional associative memory~(QiBAM)~\cite{sarkar2021qibam}, specifically to address the requirements for approximate alignment of DNA sequences.
We also proposed the quantum accelerated sequence reconstruction~(QuASeR)~\cite{sarkar2021quaser} strategy to perform de novo assembly.
This is formulated as a QUBO and solved using QAOA on a gate-model simulator, as well as, on a quantum annealer.
Quantum-accelerated genome sequence reconstruction was not studied before.
However, it was found that, this dearth of previous work is not because it is not suited for quantum acceleration.
The field of quantum computing is young and under rapid development.
This research was among the first few to explore the applications of quantum computing in bioinformatics.

There exists a considerable technological readiness gap between the resource requirements of realistic quantum algorithms and available hardware quantum processors.
Thus, most developments in QC algorithms are agnostics to hardware developments and focuses on theoretical proofs of advantages in terms of computational resources.
On the other end of the spectrum, there are trivial algorithms that are used to demonstrate and benchmark the computing capabilities of quantum hardware platforms and characterizing the instrumentation.
The intention of this research instead was to implement a proof of concept simulation of application driven quantum algorithm, thereby, to decompose the mathematical formulations and oracles in terms of quantum logic gates that can be executed on a classical quantum computing simulator using the PISQ appraoch.

%% file: 05-PathForward.tex
\section{The New Path Forward?}

\begin{figure}[bht!]
    \centering
    \includegraphics[width=0.8\linewidth]{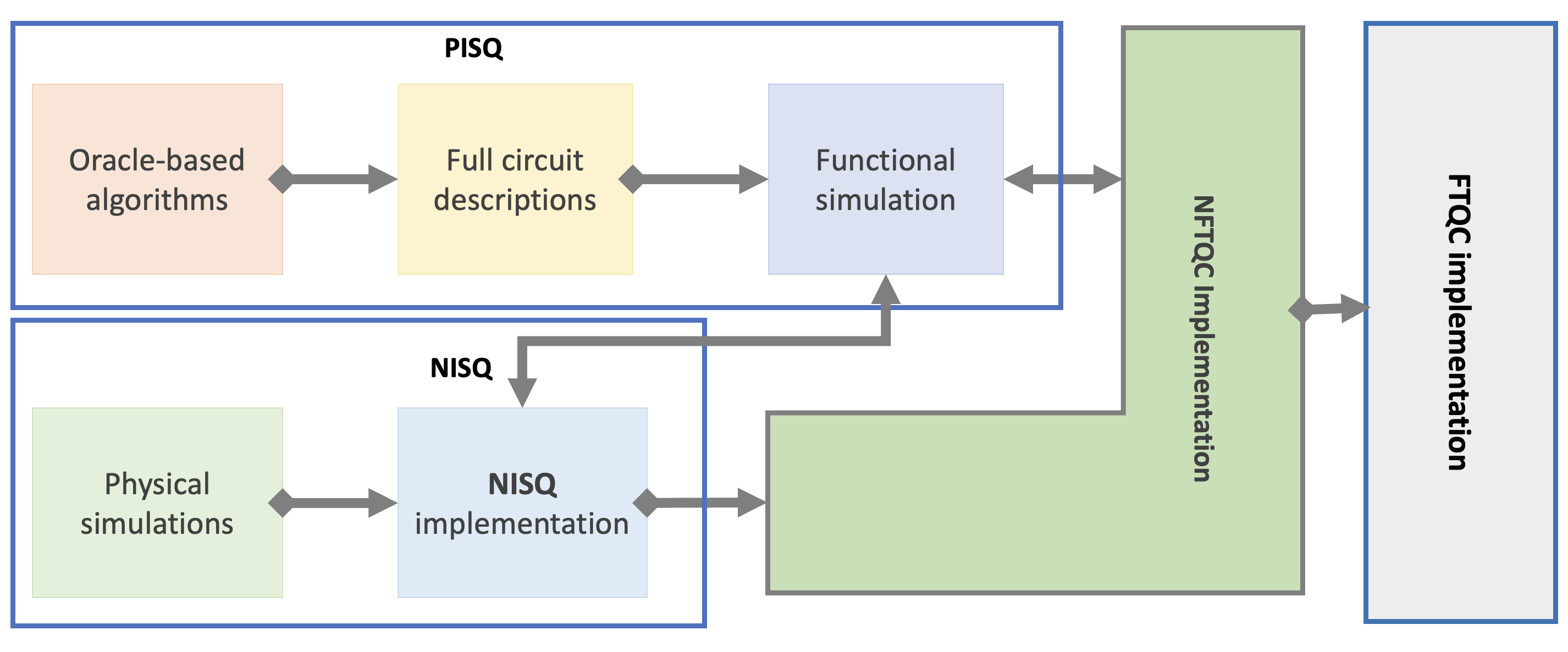}
    \caption{QC Overall Structure}
    \label{Time_Frame_Topics}
\end{figure}
It is very clear that the quantum computing R\&D effort is understandably dominated by the physicists, mathematicians and electronic engineers.  However, we argued that researchers from all scientific fields, ranging from computer science to space and chemical engineering need  to adapt an important and fundamental direction in their research and development. This will have a substantial impact on all the scientific and even economic efforts world-wide.  We therefore ask all scientists  to \textbf{explicitly} and vigorously embrace also the PISQ-oriented research line. \textbf{PISQ} stands for \textbf{Perfect Intermediate-Scale Quantum} computing and is a complementary approach to NISQ, The PISQ approach provides an abstraction layer, so no direct reliance on the roadblocks and/or progress of the quantum hardware-based  chip development efforts.

Whilst most quantum computing frameworks provide this approach, it often has as a second-hand citizen role. The perfect simulator is typically not advocated in the final product development to encourage more widespread adoption of one specific NISQ-era platform. In the long term, Fault-Tolerant Quantum Computing will allow us to integrate NISQ and PISQ results.  

We will list again the expected \textbf{ results} for the coming years in the quantum hardware and software worlds.

\begin{itemize}
    \item \textbf{Quantum hardware - }The quantum hardware physicists working on any qubit technology will have to find ways to solve all the challenges we defined in Section \ref{QChallenges}. This will also have repercussions on the kind of micro-architecture that is needed. 
    \item \textbf{Number and quality of  hardware qubits - } When developing new quantum gates, the quality of the qubits will certainly improve. This is what the evolution of the error rates represent. Hopefully, the scalability will also get better such that we can use more qubits in the quantum applications.
    \item  \textbf{Small research teams results - } As stated, small research groups are among the best in the world. Their results could be bought by some of the large players, who are focusing on the large number of qubits. With all the errors they still have.

\end{itemize}
The software part should also be initiated in parallel with the hardware development. That is the part of the industrial market that will be the end-users of whatever quantum device will reach the market.  As described in this paper, some advantages can be expected.
\begin{itemize}
    
    \item \textbf{Industrial fields - } Certain industrial fields are already looking at the quantum implications for their domain. We have started working on Earth Observation, quantum genetics and quantum computational chemistry.  Especially in the last two fields, we have already achieved good results but a lot of international collaboration is still needed.  This is where industrial companies and universities can collaborate much more.
    \item \textbf{Development of new gates and methods - } Our experience and those of others, have shown that developing completely new quantum algorithms, also allows to define, develop and use new quantum gates. Those gates can be supported either by a compiler that translates them to the universal quantum gate set. However, it could also result in new hardware, with serious implications to the micro-architecture. A similar observation can be done for quantum methods that can be re-used for other applications.
    \item \textbf{Use of up to 70-80 perfect qubits - } The PISQ-methodology allows to focus on the functional correctness of new quantum circuits, but not be restricted to a low number of hardware-based  qubits. Using perfect qubits, one can go up to 70+ qubits on supercomputers. This is still a relatively low number but we cannot expect that one approach or solution will solve all the problems.
    \item \textbf{Quantum validation and verification - } An important aspect of any kind of software development is its validation and verification. Several research groups in Europe are looking at those aspects. 
\end{itemize}

%% file: 06-Conclusion.tex
\section{Conclusion}

In the PISQ-methodology, we focus exclusively on the notion and use of perfect qubits, meaning that the qubits are now purely virtual and not connected to any hardware qubit implementation. The approach has a broader reach, in the sense that more adopters from any scientific field will step towards research for quantum computing. That also means that application developers do not have to worry, for example, about decoherence and quantum errors in the operations but rather focus on the quantum logic for important problems of their interest.

This approach is the maturation of a practical but also long-term thought process. Our multiple years of past collaboration in Delft,  the quantum computer architecture team focused on both semiconducting and superconducting qubits. During the last 10+ years we learned about the limitations of using purely physical hardware qubits, which we pointed out in this article. Still a lot of improvements has to be realised to arrive at stable and high-quality qubit implementations. 

Quantum applications formulated using perfect qubits can be executed and tested on quantum simulators running on powerful classical computers. This way, we can study and analyse new quantum algorithms for any kind of complex problems. Of course, the long-term goal is that we can make a relatively simple move to any hardware-based quantum computer or accelerator such that modern world problems can be analysed and solved. To this line of research, we invite all researchers wherever in the world they are to, such that collectively we all harvest the incredible opportunities for substantial earth improvements.

